\documentclass[twocolumn,showpacs,floats,floatfix,superscriptaddress,aps,pra]{revtex4-1}
\usepackage{amsfonts}
\usepackage{amssymb} 
\usepackage{amsmath}
\usepackage{calc}
\usepackage{dcolumn}
\usepackage{graphicx}
\usepackage{bm}
\usepackage{epstopdf}
\usepackage{epsfig}
\usepackage{float}

\usepackage {color}
\usepackage{textcomp}
\usepackage{soul}

\begin{document}

\title{Ultrabroadband beam splitting in a dissipative system of three waveguides}

\author{Rim Alrifai}
\email[]{rim.alrifai@univ-lorraine.fr}

\affiliation{Universit\'e de Lorraine, CentraleSup\'elec, LMOPS, F-57000 Metz, France}

\author{Virginie Coda}
\affiliation{Universit\'e de Lorraine, CentraleSup\'elec, LMOPS, F-57000 Metz, France}

\author{Jonathan Peltier}
\affiliation{Universit\'e de Lorraine, CentraleSup\'elec, LMOPS, F-57000 Metz, France}

\author{Andon A. Rangelov}
\affiliation{Department of Physics, Sofia University, James Bourchier 5 Boulevard, 1164 Sofia, Bulgaria}

\author{Germano Montemezzani}
\affiliation{Universit\'e de Lorraine, CentraleSup\'elec, LMOPS, F-57000 Metz, France}

\begin{abstract}
We show that a system of three parallel waveguides, among which the central one is dissipative, leads to an ultrabroadband power splitting associated with an overall 50\% power loss. The present approach is reminiscent of non-Hermitian systems in quantum mechanics and does not require a perfect effective index matching between the external and the central waveguides. The present concept does not need any slow adiabatic evolution of the waveguide parameters and may therefore be realized over very short device lengths, especially in the case where the central waveguide is of the plasmonic type. \end{abstract}

\maketitle


\section{Introduction}

Integrated optics relies on beam splitters as a crucial tool. Such an element splits the input optical signal into two or more outputs, each carrying ideally equal power. While in some cases such devices are required to work at a specific light wavelength, in general disposing of achromatic structures with a flat spectral response is very important. Today, the most commonly used integrated 1-to-2 power splitters are Y-junction and directional couplers. The former can be in principle optimized over an extended wavelength range but require very careful design and fabrication \cite{zhang2013,Lin2019}. Directional couplers \cite{Book_Ebeling_1993} rely on evanescent coupling and are easier to fabricate, but their response is highly dispersive so that a target operation wavelength must be exactly specified. The demand for integrated broadband beam splitters is particularly required in silicon-on-insulator platforms for power splitters, microring resonators, Mach-Zehnder interferometers, and polarization filters, especially those operating in a wide range of wavelengths.  Wavelength-insensitive beam splitters have been demonstrated in recent years using multimode interference couplers, subwavelength gratings directional couplers, curved directional couplers, and adiabatic couplers \cite{chen2017,zhang2020}.

Indeed, approaches based on spatially and adiabatically evolving configurations have emerged as alternative robust solutions for waveguide-based broadband and design-tolerant functionalities, such as light transfer and wave splitting. Specifically, optical analogous to quantum mechanical adiabatic processes, such as stimulated Raman adiabatic passage (STIRAP) \cite{vitanov_stimulated_2017, paspalakis_adiabatic_2006,longhi_quantum-optical_2009} have attracted much attention in this context. These quantum-like implementations for achromatic guided-wave beam splitting include the use of the so-called fractional STIRAP process \cite{dreisow_polychromatic_2009}, multiple beam splitting in a non planar configuration reminiscent of tripod STIRAP \cite{Rangelov-tripod_2012}, 1-to-$n$ multiple beam splitting exploiting the robust evolution of the adiabatic transfer state in specifically evolving planar arrangement of multiple waveguides \cite{Ciret_2012}, polarizing beam splitters in adiabatic systems of three anisotropic waveguides \cite{AlRifai_PRA19}, analogs to the two-state STIRAP process in gradually detuned waveguide pairs \cite{oukraou_control_2017}, and other interesting light redistribution functionalities involving various detuned three-waveguide configurations \cite{Hristova_PRA16}. The above adiabatic beam-splitting solutions require a slow evolution and therefore the total propagation distance cannot be shorter than a certain minimum value. Recently accelerated approaches based on shortcuts to adiabacity \cite{Chen_2018,Chung_2019} or dressed-state methods \cite{Dou_2020} have been proposed in order to reduce the required footprints, at the expense of weaker robustness and tolerances. 

In this paper, we describe extremely robust ultrabroadband beam splitting without the need for any adiabatic or superadiabatic evolution of the waveguide's parameters. The present approach is related to non-Hermitian quantum mechanics \cite{Book_moiseyev_2011,Uzdin_2012} and may be directly associated to a quantum population transfer through a decaying intermediate state \cite{Vitanov_PRA97}. Its key feature is the introduction of a dissipative loss in a central straight waveguide sandwiched between two parallel lossless outer waveguides in a planar configuration. The input light into one of the external waveguides can be split into symmetric and antisymmetric modes, both involving only the two outer waveguides. Since only the even mode can couple to the lossy central waveguide, the light left over in the odd mode leads to a power splitting in the two outer channels. With the even mode being lost, the price to pay is a 3-dB overall power loss for the case where an equal power splitting in the output ports is desired. It is shown that the above mechanism is extremely achromatic. It works at any wavelength provided that the losses are big enough. Therefore, since the attenuation is related to the length of the component, very short device lengths are possible in the case where the central channel is highly lossy, as is, for instance, the case for plasmonic waveguides \cite{Fang2015,Guo2013}.

Section II presents the theoretical concept, identifies the required parameters, and illustrates the light splitting behavior with the help of coupled mode theory expressed in normalized dimensionless units. In Sec. III, we give examples calculated with the beam propagation method and discuss the expected spectral bandwidth. Finally, in Sec. IV, we discuss implementation paths and conclude.

\begin{figure}
\centering
	\includegraphics[width=\columnwidth]{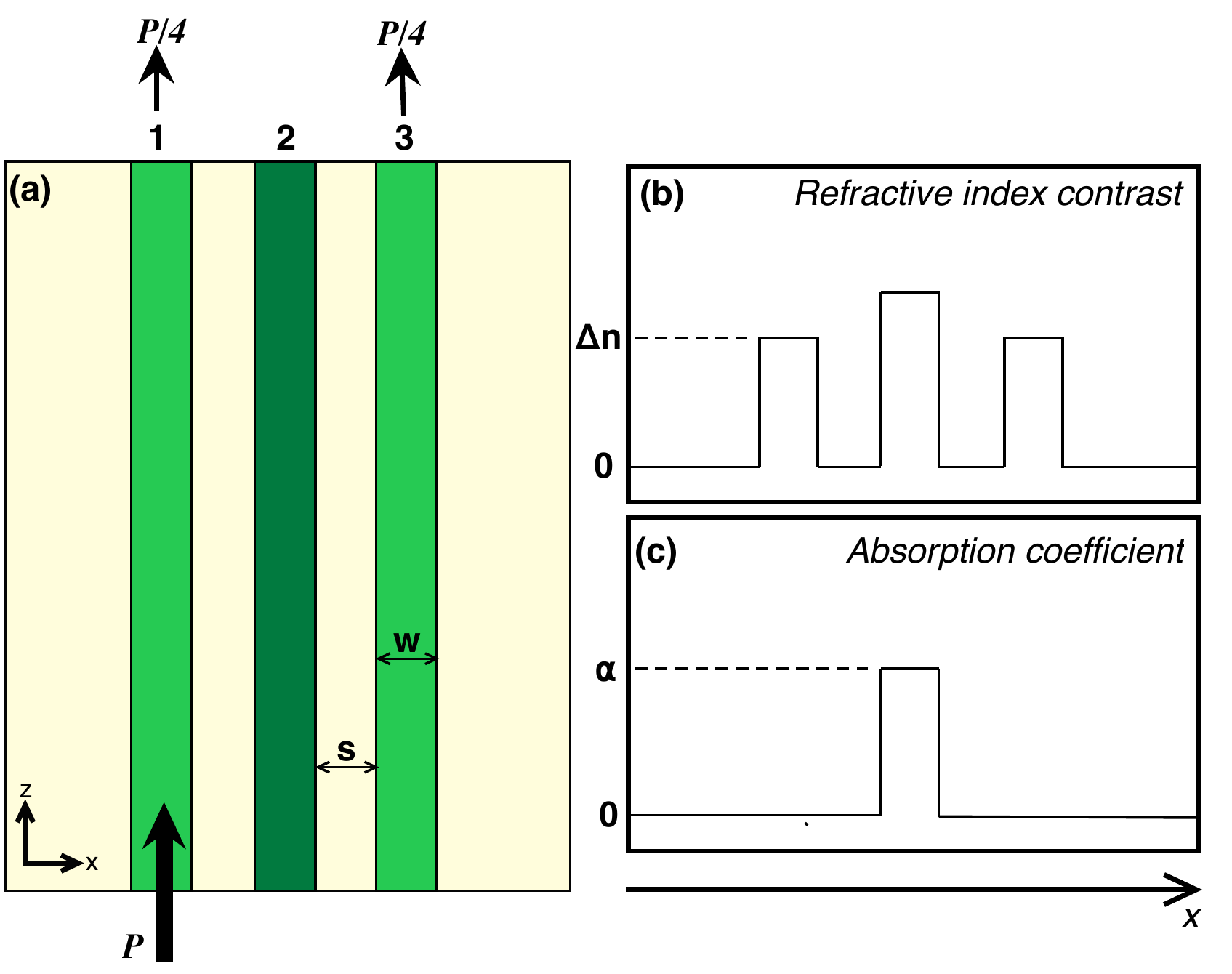}
	\caption{(a) Schematic of the three-waveguide structure based on lossy central waveguide 2 leading to beam splitting toward waveguides 1 and 3. (b) Sketches of a possible shape of the refractive index contrast $\Delta n$ as a function of the lateral dimension $x$ and (c) an absorption constant shape for the central waveguide.
Waveguides 1 and 3 are identical while waveguide 2 may differ in terms of dimensions and/or refractive index contrast (here a different  $\Delta n$ is shown). The quantities $w$ and $s$ are the core width and the core-to-core separation $s$ and are used for defining the structure for beam propagation method simulations in Sec.~\ref{Sect:Examples}.
}
\label{Fig1_configuration}
\end{figure}

\section{Theoretical background}
\label{Sect:Theory}

The key elements of the ultrabroadband beam splitter studied here are three parallel waveguides (WG) as shown in Fig. \ref{Fig1_configuration}(a). The two outer waveguides (WG1 and WG3) are supposed to be identical and lossless, while the central one (WG2) is dissipative; the related loss is expressed by the amplitude absorption constant $\alpha$. Waveguide WG2 may also differ from WG1 and WG3 in terms of geometrical dimensions and/or refractive index contrast, leading to a difference of the fundamental mode propagation constant by an amount $\Delta \beta$ with respect to the one for the outer WGs. We suppose that evanescent coupling can occur only between neighboring waveguides. The exact geometrical form of the WGs is irrelevant for the present treatment and the expressions given below hold generally independently of this form. Nevertheless, in the beam propagation calculation examples of Sec. \ref{Sect:Examples} slab-type waveguides will be used for simplicity. Under the above assumptions, in the framework of the coupled mode theory (CMT) \cite{yariv_coupled-mode_1973}, the propagation of the electric-field amplitudes $A_{1}(z)$, $A_{2}(z)$, and $A_{3}(z)$ of the waves traveling in the three evanescently coupled WGs is described by a system of three coupled differential equations. The latter can be brought in the following symmetrized matrix form (see, for instance, the transition from Eq.~(2) to Eq.~(3) in Ref.~\cite{Ciret_EIT_2013}),
\begin{equation}
i\frac{d}{dz}\left[
\begin{array}{c}
A_{1} \\
A_{2} \\
A_{3}%
\end{array}%
\right] =\left[
\begin{array}{ccc}
0 &C_{P} & 0 \\
C_{P} &\Delta \beta -i\alpha & C_{S} \\
0 & C_{S} & 0%
\end{array}%
\right] \left[
\begin{array}{c}
A_{1} \\
A_{2} \\
A_{3}%
\end{array}%
\right] ,  \label{three-state atom-row}
\end{equation}%
where  $C_P \equiv \sqrt{C _{12}C _{21}}$ and $C_S \equiv \sqrt{C _{23}C _{32}}$ are the effective coupling coefficients between WG1 and WG2 and between WG2 and WG3, respectively. Here $C_{ij}$ is the coupling constant of WG $i$ to WG $j$ and the subscripts $P$ and $S$ mean "pump" and "Stokes", in analogy with the pump and Stokes pulses used in the conventional quantum STIRAP process~\cite{vitanov_stimulated_2017}. For the symmetric situation depicted in Fig. \ref{Fig1_configuration}, one has $C_P=C_S$  but the above expression is valid in general also if this symmetry is broken. 

For our purposes, it is convenient to transform the state vector $\vec{A}=(A_1,A_2,A_3)$ into a new state vector $\vec{B}=(B_b,B_2,B_d)$  by means of following transformation:
\begin{equation}
\left[
\begin{array}{c}
B_{b} \\
B_{2} \\
B_{d}%
\end{array}%
\right] =\left[
\begin{array}{ccc}
\sin\theta &0 & \cos\theta \\
0 &1 &0 \\
\cos\theta & 0 & -\sin\theta%
\end{array}%
\right] \left[
\begin{array}{c}
A_{1} \\
A_{2} \\
A_{3}%
\end{array}%
\right] ,  \label{eq-transformation}
\end{equation}%
where $B_{b}$ and $B_{d}$ take the role of the so-called bright and dark states of quantum population dynamics \cite{Vitanov_PRA97}, respectively, and $B_{2}\equiv A_{2}$. The angle $\theta$ is defined as $\tan\theta=C_{P}/C_{S}$.
In the new coordinates, Eq.~(\ref{three-state atom-row}) takes an interesting form
\begin{equation}
i\frac{d}{dz}\left[
\begin{array}{c}
B_{b} \\
B_{2} \\
B_{d}%
\end{array}%
\right] =\left[
\begin{array}{ccc}
0 &C_{0} & 0 \\
C_{0} &\Delta \beta-i\alpha & 0 \\
0 & 0 & 0%
\end{array}%
\right] \left[
\begin{array}{c}
B_{b} \\
B_{2} \\
B_{d}%
\end{array}%
\right] ,  \label{Hamiltonian-bright-dark}
\end{equation}%
where $C_{0} \equiv \sqrt{C_{P}^2 + C_{S}^2}$. The above equation clearly shows that the dark state $B_d$ is completely decoupled from the other two and will conserve its initial amplitude. In contrast, the bright state $B_b$ and the state $B_2$ keep exchanging their power. Since the state $B_2$ and the bright state $B_b$ are tied, for sufficiently long propagation distances (as compared to $1/\alpha$, strong decay regime), the dissipation of state $B_2$ will ultimately kill both of them. 

The bright and dark states and the above relation~(\ref{Hamiltonian-bright-dark}) can find an easy interpretation also in a waveguide mode picture. The bright and dark modes can be interpreted as the lowest order even and odd combined modes of a system composed of only WG1 and WG3. In this context, Eq.~(\ref{Hamiltonian-bright-dark}) simply states that only the symmetric even combined mode (bright mode) associated to $B_b$ can couple to the central WG. Obviously, a coupling between the antisymmetric dark mode and WG2 is prevented due to the mutual destructive interference of the contributions coming from WG1 and WG3 to the wave in WG2.

Now consider the case where the input light is injected into WG1 only and thus $\vec{A}(z=0)=(1,0,0)$. For the symmetric case where $C_P=C_S$ ($\theta=\pi/4$), one has in the bright-dark basis $\vec{B}(z=0)=(1/\sqrt{2},0,1/\sqrt{2})$, which is an equal superposition of the bright and the dark state. As discussed above, the component in the bright state will be lost and for large distances one is left with $\vec{B}(z\rightarrow \infty)=(0,0,1/\sqrt{2})$. Back in the basis of the three WGs, this corresponds to a final state $\vec{A}(z\rightarrow \infty)=(1/2,0,-1/2)$, meaning that WG1 and WG3 will carry each 1/4 of the initial power in WG1 and the remaining 50\% of power is lost. The above splitting behavior is expected to occur universally provided that the decay is strong enough and that the waveguides are effectively coupled. In the following, we analyze the required range of parameters by means of an integration of the CMT equations (\ref{three-state atom-row}) in generalized units.

\begin{figure}
	\centering
    \includegraphics[width=\columnwidth]{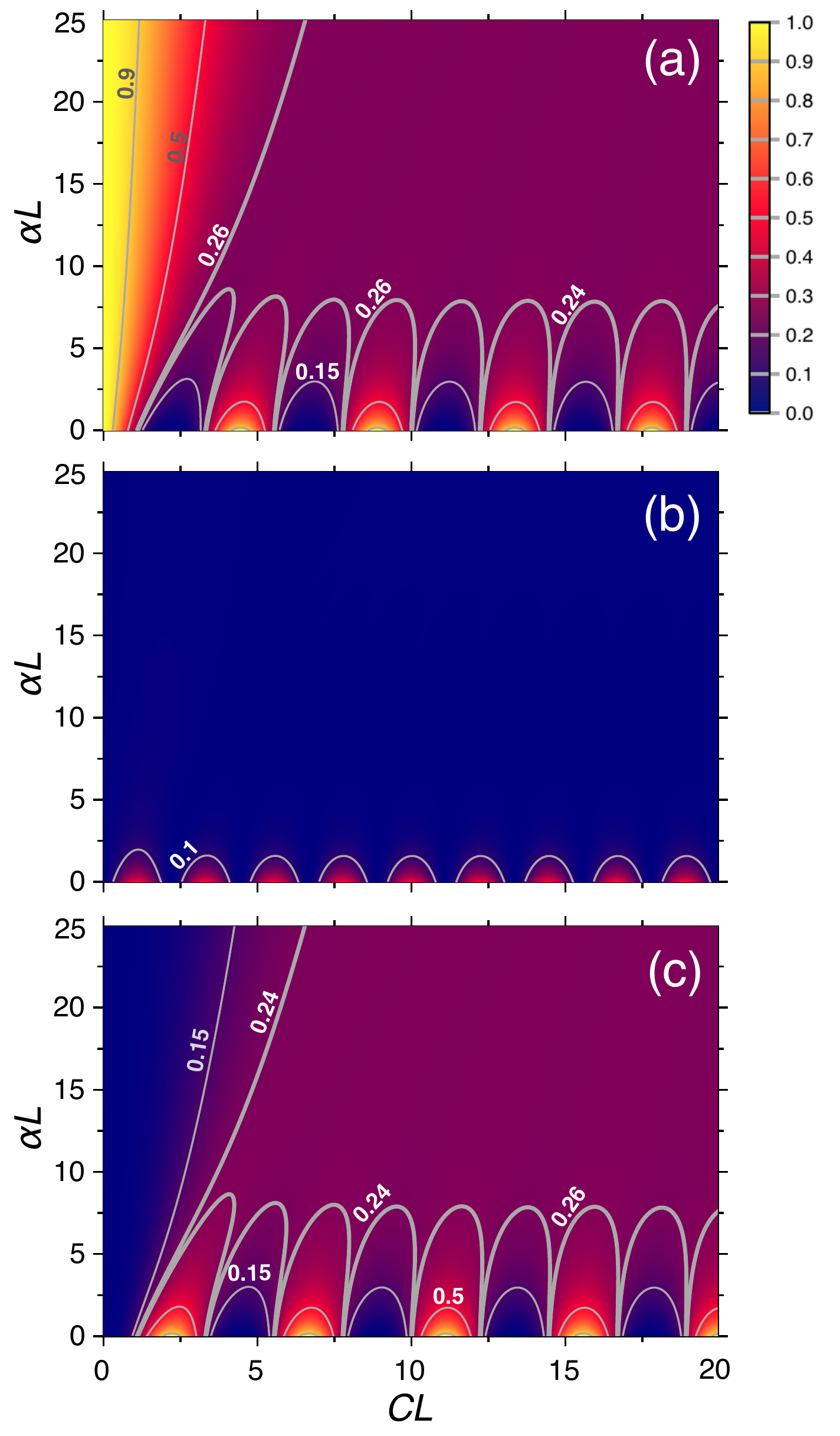}
	\caption{(a) Contour-color plot of the output power $P_1=|A_1(L)|^2$ in WG1 as obtained from an integration of Eq.~(\ref{three-state atom-row}) as a function of the products $C L$ and $\alpha L$, with $L$ being the device length. Panels (b) and (c) give the corresponding output powers $P_2=|A_2(L)|^2$ in WG2 and $P_3=|A_3(L)|^2$ in WG3, respectively. The light is input into WG1, $A_1(0)=1$. In the big top-right violet areas in panels (a) and (c), one has $P_1 \approx P_3 \approx 0.25$. Here the waveguides are resonant ($\Delta \beta = 0)$ and $C_P=C_S=C$. In order to highlight better the region where $P_1$ and $P_3$ are near 1/4, the contour lines are chosen to be not equidistant.}
	\label{Fig_color_landscape}
\end{figure}

Figure~\ref{Fig_color_landscape}
 shows the landscape of the output power in each of the three WGs upon injection of the light into WG1. The diagrams are calculated with Eq.~(\ref{three-state atom-row}) under variation of the coupling constant $C \equiv C_S = C_P$ and of the amplitude absoption constant $\alpha$. For weak values of the absorption, the light essentially oscillates between WG1 and WG3, as seen on the bottom parts of Figs.~\ref{Fig_color_landscape}(a) and  ~\ref{Fig_color_landscape}(c). These are Rabi-like oscillations equivalent to those occurring in a conventional two-waveguides directional coupler. In this regime, the system is not robust as the output depends strongly on the value of the coupling constant, which depends on the waveguide geometrical and optical parameters, as well as on the propagating wavelength. In contrast, for sufficiently large values of the products  $\alpha L$ and  $C L$ [violet upper right regions in Fig.~\ref{Fig_color_landscape}(a) and ~\ref{Fig_color_landscape}(c)], there is a vast flat land where the output power in both external WGs is close to 1/4 of the input power, as predicted by our argument above. In this region, the output distribution is virtually independent on variation of the coupling constant and the absorption constant. Figure~\ref{Fig_color_landscape} allows us to identify the boundaries of this robust region, which are roughly found for $\alpha > 5/L$ and $C>5/L$. This sets the limits for the dissipative beam splitter to work. 
 
\begin{figure*}
	\centering
    \includegraphics[width=\textwidth]{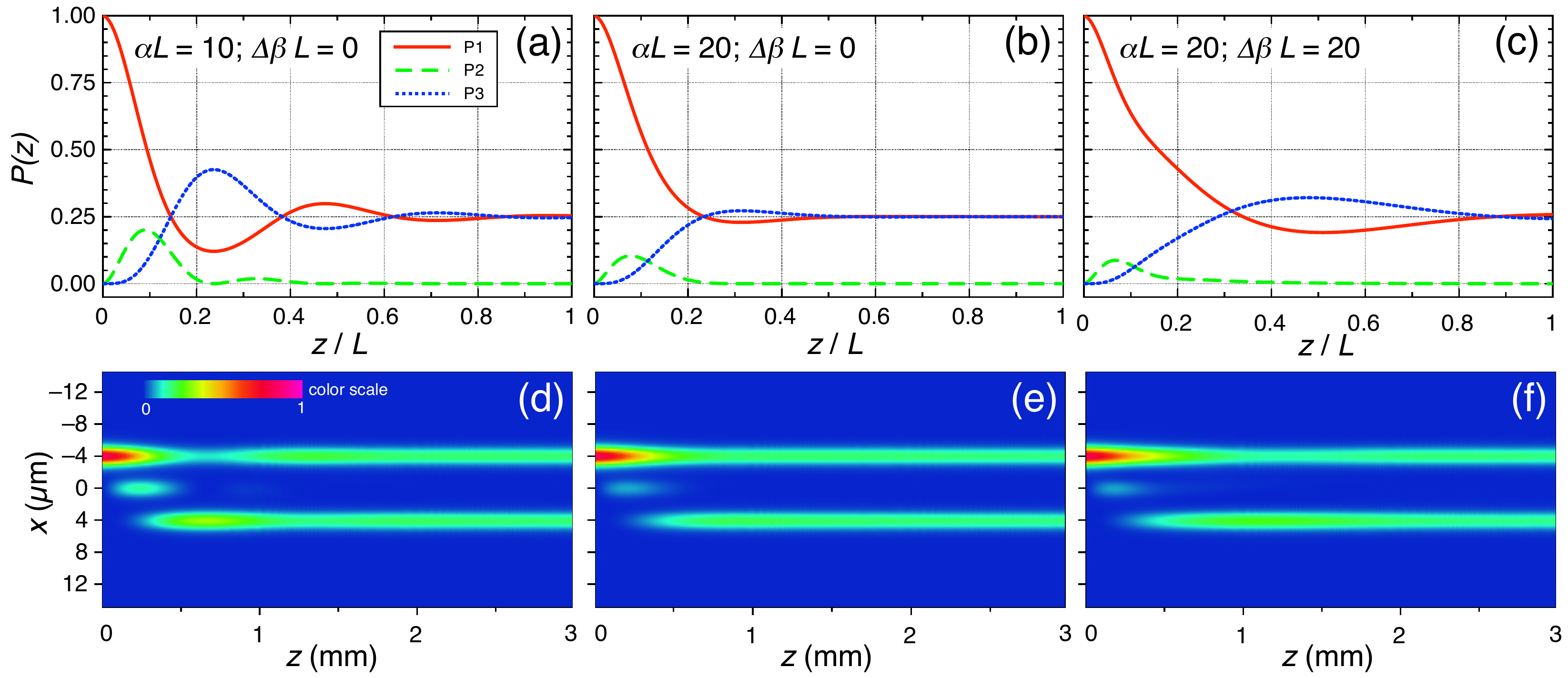}
	\caption{[(a)-(c)] Spatial evolution of the powers $P_1$ (solid red lines), $P_2$ (dashed green lines), and $P_3$ (dotted blue lines) in the three waveguides upon injection in WG1 as calculated using CMT [Eq.~(\ref{three-state atom-row})] and $C=C_P=C_S=10/L$. In panels  (a) and (b), the three waveguides are resonant (no detuning, $\Delta \beta = 0$), while in (c) WG2 is detuned by $\Delta \beta = 20/L$. The absorption constants associated to WG2 are given in the inset.  [(e)-(f)]: Light wave evolution calculated using the BPM method for equivalent structures to the cases (a)-(c), respectively. Here $L $= 3 mm, $s=w=2 \mu$m, and $\lambda=1550$ nm. In panels (d) and (e), $\Delta n_1 = \Delta n_2 = \Delta n_3 = 0.051$ in all waveguides. In panel (f), the refractive index contrast of the central waveguide is detuned with respect to the outer ones, $\Delta n_1 = \Delta n_3 = 0.051 - \delta$ and $\Delta n_2 = 0.051 + \delta$, with $\delta=9.66 \times 10^{-4}$. The absorption is limited to the core region of WG2 and equals $\alpha = 4.17$ mm$^{-1}$ in panel (d) and $\alpha = 8.33$ mm$^{-1}$ in panels (e) and (f).
	}
	\label{Fig_P(z)-and-BPM}
\end{figure*}

The longitudinal spatial evolution of the power in the three waveguides is depicted in Figs.~\ref{Fig_P(z)-and-BPM}(a) and ~\ref{Fig_P(z)-and-BPM}(b) for the case where the three waveguides are resonant and their fundamental modes possess the same propagation constant ($\Delta \beta = 0$). As seen in Fig.~\ref{Fig_P(z)-and-BPM}(a) for an amplitude absorption constant $\alpha = 10/L$, the behavior is characterized by damped oscillations of the power between the two external waveguides before reaching the stationary output state with $P_1=P_3=1/4$. If the absorption constant is doubled [Fig.~\ref{Fig_P(z)-and-BPM}(b)], the oscillations are more strongly damped and the stationary state is reached earlier. Indeed, this stationary state is reached in both cases after a distance of about 10 times the amplitude absorption length $1/\alpha$. We have verified that for further increase of $\alpha$ the above transient oscillations disappear completely and the spatial dynamics becomes overdamped, as is the case for equivalent systems in classical or quantum physics \cite{Shore_2006}.

We have already mentioned the remarkable feature that a matching of the propagation constant $\beta$ of all waveguides is not required for the beam splitting to occur. This is shown in Fig.~\ref{Fig_P(z)-and-BPM}(c), where the central waveguide is detuned by an amount $\Delta \beta = 20/L$ with respect to the two outer ones. It can be clearly recognized that the final state $\vec{A}(z\rightarrow \infty)$ does not change despite the fact that the waveguides are no longer resonant. Nevertheless, the transient spatial oscillations become slower. Indeed, for a given value of the coupling constant $C$ and of $\alpha$, we have observed that an increase of $\Delta \beta$ leads to a stronger oscillatory character. Therefore, for very large $\Delta \beta$ (as compared to $C$ and $\alpha$), a longer distance is required before reaching the stationary final state.

\section{Examples, spectral behavior}
\label{Sect:Examples}
 
The behavior discussed above can be applied to a large number of different waveguide structures of different total length $L$ as it is sufficient to fulfill the conditions for the coupling constant $C$ and the absorption constant $\alpha$ as referred to the reciprocal length $1/L$. In the following, we give concrete examples in the case of some slab waveguide structures as calculated by a split-step Fourier beam propagation method (BPM) \cite{VanRoey_81,Book_Kawano} without further assumptions. Referring to Fig.~\ref{Fig1_configuration}, we choose a structure with $s=w=$ 2 $\mu$m and $\Delta n = 0.051$, as well as a total propagation distance of 3 mm. The absorption is assumed to be limited to the core region of WG2 only. The above parameters fulfill well the conditions used to draw the first line in Fig.~\ref{Fig_P(z)-and-BPM} by the CMT calculations if one chooses a wavelength of 1550 nm and the background (cladding) refractive index $n=1.444$ of fused silica at this wavelength \cite{Malitson65}. Figure~\ref{Fig_P(z)-and-BPM}(d) gives the wave propagation in such a structure for $\alpha=4.17$ mm$^{-1}$ and shows clearly that the behavior expected from Fig.~\ref{Fig_P(z)-and-BPM}(a) is well reproduced. The same is true for a comparison of Fig.~\ref{Fig_P(z)-and-BPM}(e) with Fig.~\ref{Fig_P(z)-and-BPM}(b), where the absorption constant has been doubled. We note here that the values of  $\alpha$ used for the BPM simulations are slightly higher than those used for the CMT calculations ($\alpha L = 12.5$, respectively 25, instead of $\alpha L = 10$, respectively 20) in order to take into account the fact that the part of the wave in WG2 in the cladding is not absorbed. On the same line, even though this cannot be recognized directly in Figs.~\ref{Fig_P(z)-and-BPM}(d) and ~\ref{Fig_P(z)-and-BPM}(e), we would like to mention that the BPM simulations show a weak dissipation of the power in WG1 and WG3 in the stationary state toward the end of the propagation so that the output power in each of the outer WGs is slightly less than 23\% in Fig.~\ref{Fig_P(z)-and-BPM}(d) rather than the theoretically expected 25\%. This is related to the fact that the waveguide modes of WG1 and WG3 still touch the core region of WG2 and are weakly absorbed by this effect.  The associated effective absorption constant is less than 0.3\% as compared to the value of  $\alpha$ in the core of WG2. Nevertheless, to avoid an excessive influence of this effect, it is important to limit the length of the device to about 10 absorption lengths, as discussed already above.  Finally, Fig.~\ref{Fig_P(z)-and-BPM}(f) shows the expected propagation in the nonresonant case, where the central waveguide differs from the outer ones. This BPM simulation confirms that the final wave splitting is reached despite for the waveguide detuning, as discussed in Sec.~\ref{Sect:Theory}.
 
\begin{figure}
	\centering
    \includegraphics[width=\columnwidth]{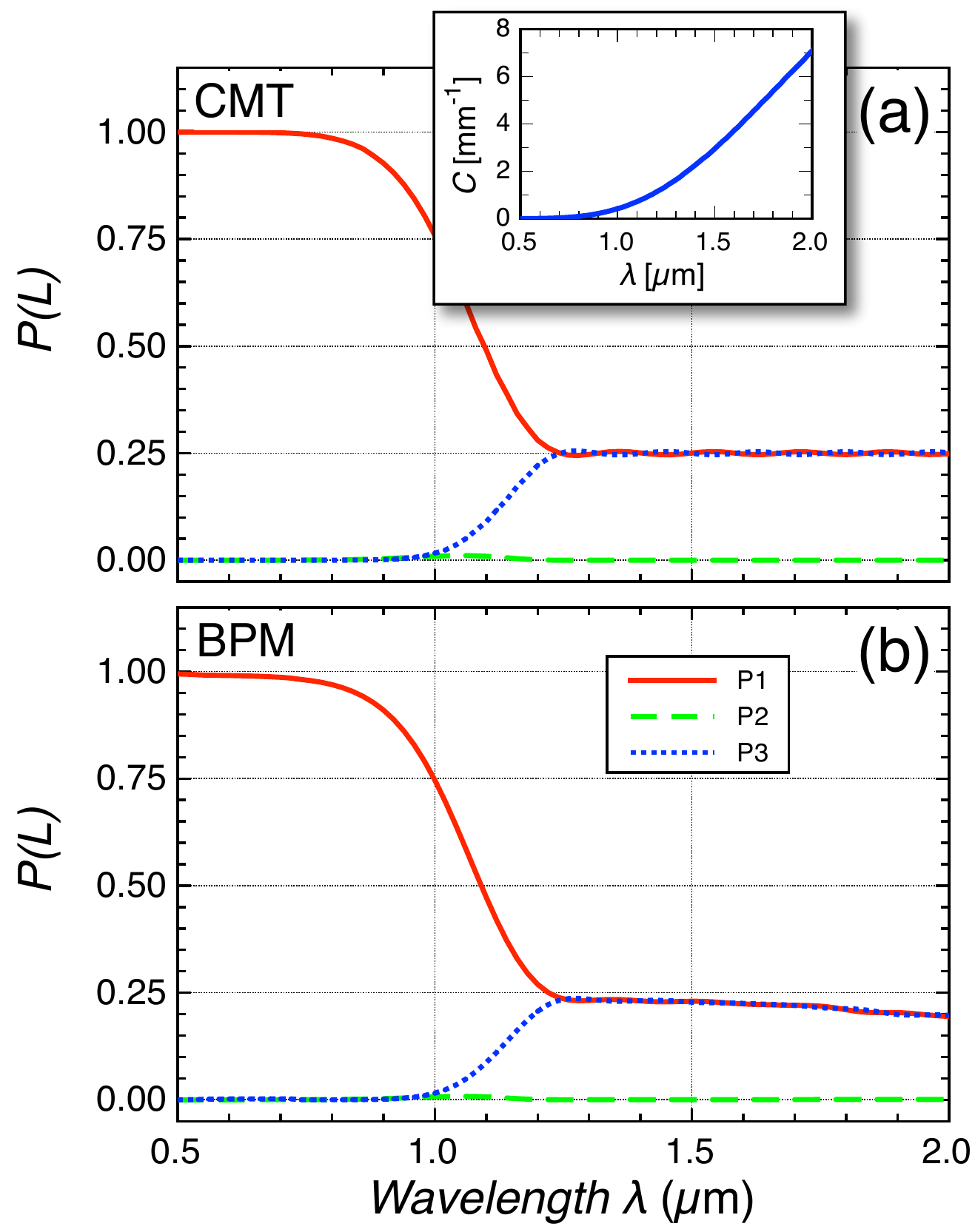}
	\caption{Output power in each waveguide as a function of wavelength as calculated by (a) CMT and (b) BPM. The waveguide structure is the same as the one used for Fig.~\ref{Fig_P(z)-and-BPM}(d) with a dispersion of the background fused silica refractive index according to Ref. \cite{Malitson65}. The coupling constant $C$ depends on wavelength according to the inset in the upper panel. The absorption in WG2 is assumed constant over the whole spectrum with $\alpha L=10$ for CMT and $\alpha = 4.17$ mm$^{-1}$ in the core of WG2 for BPM.}
	\label{Fig_wavelength}
\end{figure}

Next we discuss the wavelength dependence of the output power distribution among the parallel waveguides. As discussed above, it is expected that the outer waveguides carry both 1/4 of the input power provided that the products $C L$ and $\alpha L$ are large enough. Nevertheless, for a given waveguide structure, a decrease of the wavelength leads to a decrease of the coupling constant $C$ because the modes are better confined. The consequence is that for a given waveguide structure there is a minimum wavelength for which the beam splitting still works. Indeed, for too short wavelengths, the light remains essentially confined in the input waveguide. This behavior is depicted in Fig.~\ref{Fig_wavelength} for the same resonant structure used to simulate Fig.~\ref{Fig_P(z)-and-BPM}(d) for fused silica waveguides. Here the onset wavelength is at about 1.25 $\mu$m both for the CMT and the BPM calculations. This value can be easily shifted to shorter or longer wavelengths by modifying the waveguide structure in terms of the waveguide widths, distances, and refractive index contrast. Note also that the slight decrease of the powers $P_1$ and $P_3$ at the output of WG1 and WG3 seen in the BPM calculation for the longer wavelengths is due to the same reasons discussed above. At longer wavelengths, the modes in WG1 and WG3 are less tightly confined and overlap more with the core of WG2, where part of the light is being absorbed. This effect is not included in the CMT calculation.

\begin{figure}
	\centering
    \includegraphics[width=\columnwidth]{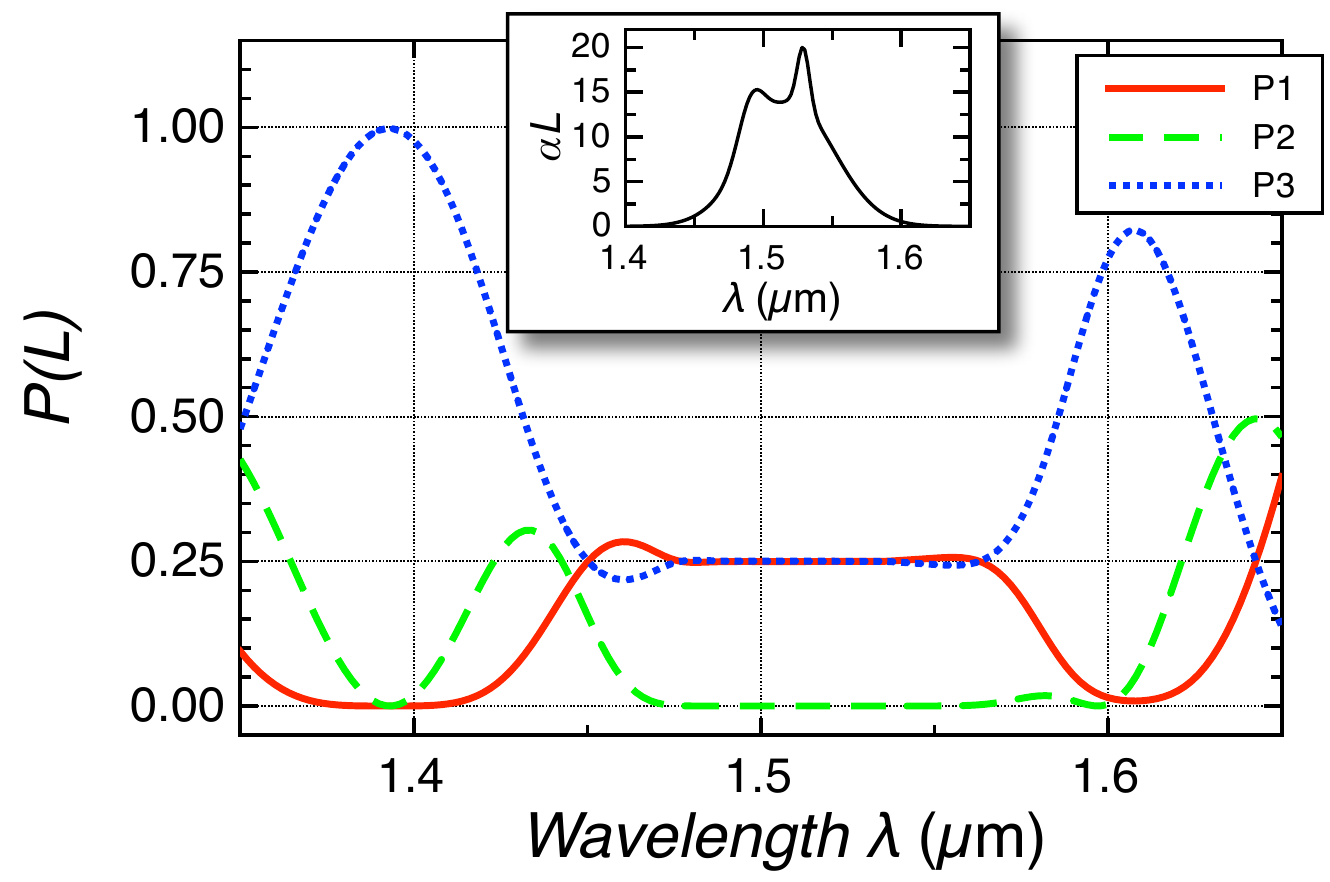}
	\caption{CMT prediction for the output power in WG1 (solid red line), WG2 (green dashed line), and WG3 (dotted blue line) in the case where the absorption constant in WG2 corresponds to the Er$^{3+}$ spectrum extracted from Ref. \cite{Naresh2015} given in the inset.}
	\label{Fig_wavelength-Er}
\end{figure}

In the case where the absorption is limited to a region of the spectrum, the dissipative beam splitter will act essentially over the spectral width of this absorption region. This is illustrated in Fig.~\ref{Fig_wavelength-Er} where the typical absorption spectrum of Er$^{3+}$ ions \cite{Naresh2015,Tikhomirov2002, Rolli2003} with maximum near $\lambda = 1.53$ $\mu$m is assumed with a peak value at $\alpha L=20$. The output powers are calculated here by the CMT method for the same structure as in Figs.~\ref{Fig_P(z)-and-BPM}(d) and~\ref{Fig_wavelength} but for the case of a wavelength-dependent loss in WG2.  It can be seen that in this specific case the broadband spectral region where the structure leads to achromatic beam splitting into output ports 1 and 3 extends roughly from 1.47 to 1.56 $\mu$m, corresponding to the region for which the product $\alpha L \geq 5$. A further increase of the operation spectral width using the same dopant requires an increase of the peak value of $\alpha L$.

\section{Discussion and Conclusion}

One possibility to realize the required dissipative waveguide is by intentionally doping with an absorbing ion for the desired wavelength range. In the case of Er$^{3+}$, illustrated above, doped waveguides can be realized, for instance, in LiNbO$_3$ \cite{Sohler2005, Mattarelli2006}, glasses \cite{Yan1997, Wetter2019}, or polymers \cite{Sloof2002,Ziarko2019}. In all these cases, achievable doping levels lead to absorption constants in the order of few cm$^{-1}$ so that the required device length will be on the order of centimeters. The same order of magnitude can be expected for the doping with other rare earth ions such as Ho$^{3+}$, Tm$^{3+}$, or Nd$^{3+}$, or for the doping with metal impurities such as Fe, Cu, Ni or others. A certain increase of the dissipation losses and thus shorter devices may be expected by involving scattering processes in addition to absorption, for instance, by the incorporation of nanometer-size particles or crystals \cite{Wetter2019,Catchpole2006} into the waveguides. Moreover, as an alternative one may also rely exclusively on scattering losses by controlling waveguide imperfections. This was used for instance in a recent fundamental study on PT-symmetric systems by inducing controlled dwelling scattering centers in laser-written fused silica waveguides \cite{BiesenthalPRL2019}. The achieved intensity dissipation constant of the order of 1 cm$^{-1}$ would require device lengths of several centimeters for this specific approach.

A better alternative to impurity doping and/or controlled scattering consists in combining the two outer dielectric waveguides with a central plasmonic waveguide. This approach takes advantage of the strong metallic losses and of their weak wavelength specificity to allow for shorter devices with an ultrabroad spectral operation range. The efficient evanescent coupling between dielectric and plasmonic waveguides can be realized in a standard way with today's silicon photonics \cite{Li2010,Briggs2010,Lou2012, Chee2012,Guan2013}. Typically, coupling lengths of the order of few micrometers are being realized, so that achieving a product $CL \approx 5$ for a length $L$ of roughly 10 $\mu$m is feasible. The intensity losses for plasmonic waveguides can vary strongly according to the specific geometry. Unlike for other plasmonic applications, since our concept relies on these losses, here an optimization of the propagation distance of the plasmon-polariton is not required and is even detrimental if a short device footprint is wished. Plasmon propagation distances $L_p \equiv 1/(2 \alpha)$ of the order of few micrometers are quite usual in different types of plasmonic waveguides. If we assume $L_p = 5$ $\mu$m, the required product $ \alpha L > 5$ is reached for a device length of 50 $\mu$m, which is by far shorter than what can be achieved for broadband solutions based on adiabatically evolving waveguide systems, such as STIPAP-like configurations.

Finally, we would like to mention that the present approach is not limited to the case where the two output ports (WG1 and WG3) carry the same power. The power ratio between WG1 and WG3 can be varied by modifying the relative strengths of the coupling constants $C_P$ and $C_S$, for instance, by having asymmetric distances for WG1 and WG3 from the central WG. A modification of the ratio $C_P/C_S$ leads to a different contribution of the individual WG states $A_1$ and $A_3$ to the surviving dark state $B_d$. Note that for an asymmetric system it is advantageous to choose $C_S > C_P$ and thus a smaller value of the angle $\theta$ so that the contribution of the dark state to the initial input state $A_1$ is maximized. For instance, for $C_S = 2 C_P$ at the output WG1 will carry four times more power than WG3 and only 1/5 of the total input power will have been dissipated in the process. In the reverse case ($C_P = 2 C_S$), the output power in WG3 will be four times the one of WG1 but 4/5 of the total input power will have been lost. 

In conclusion, we have presented a concept for an ultrabroadband integrated power divider where losses within the system are turned to an advantage. The system consists of a strongly lossy waveguide sandwiched between two other (lossless) waveguides all aligned in the same plane.  The proposed device does not require perfect phase matching between the outer and the central waveguides and works properly as long as the products $C L$ and $\alpha L$ both exceed roughly the value of 5 over the wavelength range of interest. It may be implemented by different approaches including the selective doping of the central waveguide, the controlled inclusion of scattering centers into it, or the use of a plasmonic-type central waveguide. In the latter case, device lengths of few tens of micrometers can be envisaged. For an equal power division into the two output ports, a fundamental overall total power loss of 3 dB must be paid, which can be reduced in the case of asymmetric power division.

\section*{Achnowledgements}

Part of this work was supported by Sofia University Grant No. 80-10-191/2020.

\newpage


\begin{thebibliography}{44}%
	\makeatletter
	\providecommand \@ifxundefined [1]{%
		\@ifx{#1\undefined}
	}%
	\providecommand \@ifnum [1]{%
		\ifnum #1\expandafter \@firstoftwo
		\else \expandafter \@secondoftwo
		\fi
	}%
	\providecommand \@ifx [1]{%
		\ifx #1\expandafter \@firstoftwo
		\else \expandafter \@secondoftwo
		\fi
	}%
	\providecommand \natexlab [1]{#1}%
	\providecommand \enquote  [1]{``#1''}%
	\providecommand \bibnamefont  [1]{#1}%
	\providecommand \bibfnamefont [1]{#1}%
	\providecommand \citenamefont [1]{#1}%
	\providecommand \href@noop [0]{\@secondoftwo}%
	\providecommand \href [0]{\begingroup \@sanitize@url \@href}%
	\providecommand \@href[1]{\@@startlink{#1}\@@href}%
	\providecommand \@@href[1]{\endgroup#1\@@endlink}%
	\providecommand \@sanitize@url [0]{\catcode `\\12\catcode `\$12\catcode
		`\&12\catcode `\#12\catcode `\^12\catcode `\_12\catcode `\%12\relax}%
	\providecommand \@@startlink[1]{}%
	\providecommand \@@endlink[0]{}%
	\providecommand \url  [0]{\begingroup\@sanitize@url \@url }%
	\providecommand \@url [1]{\endgroup\@href {#1}{\urlprefix }}%
	\providecommand \urlprefix  [0]{URL }%
	\providecommand \Eprint [0]{\href }%
	\providecommand \doibase [0]{http://dx.doi.org/}%
	\providecommand \selectlanguage [0]{\@gobble}%
	\providecommand \bibinfo  [0]{\@secondoftwo}%
	\providecommand \bibfield  [0]{\@secondoftwo}%
	\providecommand \translation [1]{[#1]}%
	\providecommand \BibitemOpen [0]{}%
	\providecommand \bibitemStop [0]{}%
	\providecommand \bibitemNoStop [0]{.\EOS\space}%
	\providecommand \EOS [0]{\spacefactor3000\relax}%
	\providecommand \BibitemShut  [1]{\csname bibitem#1\endcsname}%
	\let\auto@bib@innerbib\@empty
	\bibitem [{\citenamefont {Zhang}\ \emph {et~al.}(2013)\citenamefont {Zhang},
		\citenamefont {Yang}, \citenamefont {Lim}, \citenamefont {Lo}, \citenamefont
		{Galland}, \citenamefont {{Baehr-Jones}},\ and\ \citenamefont
		{Hochberg}}]{zhang2013}%
	\BibitemOpen
	\bibfield  {author} {\bibinfo {author} {\bibfnamefont {Y.}~\bibnamefont
			{Zhang}}, \bibinfo {author} {\bibfnamefont {S.}~\bibnamefont {Yang}},
		\bibinfo {author} {\bibfnamefont {A.~E.-J.}\ \bibnamefont {Lim}}, \bibinfo
		{author} {\bibfnamefont {G.-Q.}\ \bibnamefont {Lo}}, \bibinfo {author}
		{\bibfnamefont {C.}~\bibnamefont {Galland}}, \bibinfo {author} {\bibfnamefont
			{T.}~\bibnamefont {{Baehr-Jones}}}, \ and\ \bibinfo {author} {\bibfnamefont
			{M.}~\bibnamefont {Hochberg}},\ }\href {\doibase 10.1364/OE.21.001310}
	{\bibfield  {journal} {\bibinfo  {journal} {Opt. Express}\ }\textbf {\bibinfo
			{volume} {21}},\ \bibinfo {pages} {1310} (\bibinfo {year}
		{2013})}\BibitemShut {NoStop}%
	\bibitem [{\citenamefont {Lin}\ and\ \citenamefont {Shi}(2019)}]{Lin2019}%
	\BibitemOpen
	\bibfield  {author} {\bibinfo {author} {\bibfnamefont {Z.}~\bibnamefont
			{Lin}}\ and\ \bibinfo {author} {\bibfnamefont {W.}~\bibnamefont {Shi}},\
	}\href {\doibase 10.1364/OE.27.014338} {\bibfield  {journal} {\bibinfo
			{journal} {Opt. Express}\ }\textbf {\bibinfo {volume} {27}},\ \bibinfo
		{pages} {14338} (\bibinfo {year} {2019})}\BibitemShut {NoStop}%
	\bibitem [{\citenamefont {Ebeling}(1993)}]{Book_Ebeling_1993}%
	\BibitemOpen
	\bibfield  {author} {\bibinfo {author} {\bibfnamefont {K.~J.}\ \bibnamefont
			{Ebeling}},\ }\href@noop {} {\emph {\bibinfo {title} {Integrated
				Optoelectronics}}}\ (\bibinfo  {publisher} {{Springer}},\ \bibinfo {address}
	{{Berlin}},\ \bibinfo {year} {1993})\BibitemShut {NoStop}%
	\bibitem [{\citenamefont {Chen}\ \emph {et~al.}(2017)\citenamefont {Chen},
		\citenamefont {Ong}, \citenamefont {Ang}, \citenamefont {Lim}, \citenamefont
		{Png},\ and\ \citenamefont {Tan}}]{chen2017}%
	\BibitemOpen
	\bibfield  {author} {\bibinfo {author} {\bibfnamefont {G.~F.~R.}\
			\bibnamefont {Chen}}, \bibinfo {author} {\bibfnamefont {J.~R.}\ \bibnamefont
			{Ong}}, \bibinfo {author} {\bibfnamefont {T.~Y.~L.}\ \bibnamefont {Ang}},
		\bibinfo {author} {\bibfnamefont {S.~T.}\ \bibnamefont {Lim}}, \bibinfo
		{author} {\bibfnamefont {C.~E.}\ \bibnamefont {Png}}, \ and\ \bibinfo
		{author} {\bibfnamefont {D.~T.~H.}\ \bibnamefont {Tan}},\ }\href {\doibase
		10.1038/s41598-017-07618-6} {\bibfield  {journal} {\bibinfo  {journal} {Sci.
				Rep.}\ }\textbf {\bibinfo {volume} {7}},\ \bibinfo {pages} {7246} (\bibinfo
		{year} {2017})}\BibitemShut {NoStop}%
	\bibitem [{\citenamefont {Zhang}\ \emph {et~al.}(2020)\citenamefont {Zhang},
		\citenamefont {Zheng}, \citenamefont {Song},
		\citenamefont {Liu}, \citenamefont {Xu},\ and\ \citenamefont
		{Majumdar}}]{zhang2020}%
	\BibitemOpen
	\bibfield  {author} {\bibinfo {author} {\bibfnamefont {F.}~\bibnamefont
			{Zhang}}, \bibinfo {author} {\bibfnamefont {J.}~\bibnamefont {Zheng}},
		\bibinfo {author} {\bibfnamefont {Y.}~\bibnamefont {Song}}, \bibinfo {author} {\bibfnamefont
			{W.}~\bibnamefont {Liu}}, \bibinfo {author} {\bibfnamefont {P.}~\bibnamefont
			{Xu}}, \ and\ \bibinfo {author}
		{\bibfnamefont {A.}~\bibnamefont {Majumdar}},\ }\href {\doibase
		10.1364/OSAC.385546} {\bibfield  {journal} {\bibinfo  {journal} {OSA
				Continuum}\ }\textbf {\bibinfo {volume} {3}},\ \bibinfo {pages} {560}
		(\bibinfo {year} {2020})}\BibitemShut {NoStop}%
	\bibitem [{\citenamefont {Vitanov}\ \emph {et~al.}(2017)\citenamefont
		{Vitanov}, \citenamefont {Rangelov}, \citenamefont {Shore},\ and\
		\citenamefont {Bergmann}}]{vitanov_stimulated_2017}%
	\BibitemOpen
	\bibfield  {author} {\bibinfo {author} {\bibfnamefont {N.~V.}\ \bibnamefont
			{Vitanov}}, \bibinfo {author} {\bibfnamefont {A.~A.}\ \bibnamefont
			{Rangelov}}, \bibinfo {author} {\bibfnamefont {B.~W.}\ \bibnamefont {Shore}},
		\ and\ \bibinfo {author} {\bibfnamefont {K.}~\bibnamefont {Bergmann}},\
	}\href {\doibase 10.1103/RevModPhys.89.015006} {\bibfield  {journal}
		{\bibinfo  {journal} {Rev. Mod. Phys.}\ }\textbf {\bibinfo {volume} {89}},\
		\bibinfo {pages} {015006} (\bibinfo {year} {2017})}\BibitemShut {NoStop}%
	\bibitem [{\citenamefont {Paspalakis}(2006)}]{paspalakis_adiabatic_2006}%
	\BibitemOpen
	\bibfield  {author} {\bibinfo {author} {\bibfnamefont {E.}~\bibnamefont
			{Paspalakis}},\ }\href {\doibase 10.1016/j.optcom.2005.07.060} {\bibfield
		{journal} {\bibinfo  {journal} {Opt. Commun.}\ }\textbf {\bibinfo {volume}
			{258}},\ \bibinfo {pages} {30} (\bibinfo {year} {2006})}\BibitemShut
	{NoStop}%
	\bibitem [{\citenamefont {Longhi}(2009)}]{longhi_quantum-optical_2009}%
	\BibitemOpen
	\bibfield  {author} {\bibinfo {author} {\bibfnamefont {S.}~\bibnamefont
			{Longhi}},\ }\href {\doibase 10.1002/lpor.200810055} {\bibfield  {journal}
		{\bibinfo  {journal} {Laser {$\&$} Photon. Rev.}\ }\textbf {\bibinfo {volume}
			{3}},\ \bibinfo {pages} {243} (\bibinfo {year} {2009})}\BibitemShut {NoStop}%
	\bibitem [{\citenamefont {Dreisow}\ \emph {et~al.}(2009)\citenamefont
		{Dreisow}, \citenamefont {Ornigotti}, \citenamefont {Szameit}, \citenamefont
		{Heinrich}, \citenamefont {Keil}, \citenamefont {Nolte}, \citenamefont
		{T{\"u}nnermann},\ and\ \citenamefont {Longhi}}]{dreisow_polychromatic_2009}%
	\BibitemOpen
	\bibfield  {author} {\bibinfo {author} {\bibfnamefont {F.}~\bibnamefont
			{Dreisow}}, \bibinfo {author} {\bibfnamefont {M.}~\bibnamefont {Ornigotti}},
		\bibinfo {author} {\bibfnamefont {A.}~\bibnamefont {Szameit}}, \bibinfo
		{author} {\bibfnamefont {M.}~\bibnamefont {Heinrich}}, \bibinfo {author}
		{\bibfnamefont {R.}~\bibnamefont {Keil}}, \bibinfo {author} {\bibfnamefont
			{S.}~\bibnamefont {Nolte}}, \bibinfo {author} {\bibfnamefont
			{A.}~\bibnamefont {T{\"u}nnermann}}, \ and\ \bibinfo {author} {\bibfnamefont
			{S.}~\bibnamefont {Longhi}},\ }\href {\doibase 10.1063/1.3279134} {\bibfield
		{journal} {\bibinfo  {journal} {Appl. Phys. Lett.}\ }\textbf {\bibinfo
			{volume} {95}},\ \bibinfo {pages} {261102} (\bibinfo {year}
		{2009})}\BibitemShut {NoStop}%
	\bibitem [{\citenamefont {Rangelov}\ and\ \citenamefont
		{Vitanov}(2012)}]{Rangelov-tripod_2012}%
	\BibitemOpen
	\bibfield  {author} {\bibinfo {author} {\bibfnamefont {A.~A.}\ \bibnamefont
			{Rangelov}}\ and\ \bibinfo {author} {\bibfnamefont {N.~V.}\ \bibnamefont
			{Vitanov}},\ }\href {\doibase 10.1103/PhysRevA.85.055803} {\bibfield
		{journal} {\bibinfo  {journal} {Phys. Rev. A}\ }\textbf {\bibinfo {volume}
			{85}},\ \bibinfo {pages} {055803} (\bibinfo {year} {2012})}\BibitemShut
	{NoStop}%
	\bibitem [{\citenamefont {Ciret}\ \emph {et~al.}(2012)\citenamefont {Ciret},
		\citenamefont {Coda}, \citenamefont {Rangelov}, \citenamefont {Neshev},\ and\
		\citenamefont {Montemezzani}}]{Ciret_2012}%
	\BibitemOpen
	\bibfield  {author} {\bibinfo {author} {\bibfnamefont {C.}~\bibnamefont
			{Ciret}}, \bibinfo {author} {\bibfnamefont {V.}~\bibnamefont {Coda}},
		\bibinfo {author} {\bibfnamefont {A.~A.}\ \bibnamefont {Rangelov}}, \bibinfo
		{author} {\bibfnamefont {D.~N.}\ \bibnamefont {Neshev}}, \ and\ \bibinfo
		{author} {\bibfnamefont {G.}~\bibnamefont {Montemezzani}},\ }\href {\doibase
		10.1364/OL.37.003789} {\bibfield  {journal} {\bibinfo  {journal} {Opt.
				Lett.}\ }\textbf {\bibinfo {volume} {37}},\ \bibinfo {pages} {3789} (\bibinfo
		{year} {2012})}\BibitemShut {NoStop}%
	\bibitem [{\citenamefont {Alrifai}\ \emph {et~al.}(2019)\citenamefont
		{Alrifai}, \citenamefont {Coda}, \citenamefont {Rangelov},\ and\
		\citenamefont {Montemezzani}}]{AlRifai_PRA19}%
	\BibitemOpen
	\bibfield  {author} {\bibinfo {author} {\bibfnamefont {R.}~\bibnamefont
			{Alrifai}}, \bibinfo {author} {\bibfnamefont {V.}~\bibnamefont {Coda}},
		\bibinfo {author} {\bibfnamefont {A.~A.}\ \bibnamefont {Rangelov}}, \ and\
		\bibinfo {author} {\bibfnamefont {G.}~\bibnamefont {Montemezzani}},\ }\href
	{\doibase 10.1103/PhysRevA.100.063841} {\bibfield  {journal} {\bibinfo
			{journal} {Phys. Rev. A}\ }\textbf {\bibinfo {volume} {100}},\ \bibinfo
		{pages} {063841} (\bibinfo {year} {2019})}\BibitemShut {NoStop}%
	\bibitem [{\citenamefont {Oukraou}\ \emph {et~al.}(2017)\citenamefont
		{Oukraou}, \citenamefont {Vittadello}, \citenamefont {Coda}, \citenamefont
		{Ciret}, \citenamefont {Alonzo}, \citenamefont {Rangelov}, \citenamefont
		{Vitanov},\ and\ \citenamefont {Montemezzani}}]{oukraou_control_2017}%
	\BibitemOpen
	\bibfield  {author} {\bibinfo {author} {\bibfnamefont {H.}~\bibnamefont
			{Oukraou}}, \bibinfo {author} {\bibfnamefont {L.}~\bibnamefont {Vittadello}},
		\bibinfo {author} {\bibfnamefont {V.}~\bibnamefont {Coda}}, \bibinfo {author}
		{\bibfnamefont {C.}~\bibnamefont {Ciret}}, \bibinfo {author} {\bibfnamefont
			{M.}~\bibnamefont {Alonzo}}, \bibinfo {author} {\bibfnamefont {A.~A.}\
			\bibnamefont {Rangelov}}, \bibinfo {author} {\bibfnamefont {N.~V.}\
			\bibnamefont {Vitanov}}, \ and\ \bibinfo {author} {\bibfnamefont
			{G.}~\bibnamefont {Montemezzani}},\ }\href {\doibase
		10.1103/PhysRevA.95.023811} {\bibfield  {journal} {\bibinfo  {journal} {Phys.
				Rev. A}\ }\textbf {\bibinfo {volume} {95}},\ \bibinfo {pages} {023811}
		(\bibinfo {year} {2017})}\BibitemShut {NoStop}%
	\bibitem [{\citenamefont {Hristova}\ \emph {et~al.}(2016)\citenamefont
		{Hristova}, \citenamefont {Rangelov}, \citenamefont {Montemezzani},\ and\
		\citenamefont {Vitanov}}]{Hristova_PRA16}%
	\BibitemOpen
	\bibfield  {author} {\bibinfo {author} {\bibfnamefont {H.~S.}\ \bibnamefont
			{Hristova}}, \bibinfo {author} {\bibfnamefont {A.~A.}\ \bibnamefont
			{Rangelov}}, \bibinfo {author} {\bibfnamefont {G.}~\bibnamefont
			{Montemezzani}}, \ and\ \bibinfo {author} {\bibfnamefont {N.~V.}\
			\bibnamefont {Vitanov}},\ }\href {\doibase 10.1103/PhysRevA.93.033802}
	{\bibfield  {journal} {\bibinfo  {journal} {Phys. Rev. A}\ }\textbf {\bibinfo
			{volume} {93}},\ \bibinfo {pages} {033802} (\bibinfo {year}
		{2016})}\BibitemShut {NoStop}%
	\bibitem [{\citenamefont {Chen}\ \emph {et~al.}(2018)\citenamefont {Chen},
		\citenamefont {Wen}, \citenamefont {Shi},\ and\ \citenamefont
		{Tseng}}]{Chen_2018}%
	\BibitemOpen
	\bibfield  {author} {\bibinfo {author} {\bibfnamefont {X.}~\bibnamefont
			{Chen}}, \bibinfo {author} {\bibfnamefont {R.-D.}\ \bibnamefont {Wen}},
		\bibinfo {author} {\bibfnamefont {J.-L.}\ \bibnamefont {Shi}}, \ and\
		\bibinfo {author} {\bibfnamefont {S.-Y.}\ \bibnamefont {Tseng}},\ }\href
	{\doibase 10.1088/2040-8986/aab02c} {\bibfield  {journal} {\bibinfo
			{journal} {Journal of Optics}\ }\textbf {\bibinfo {volume} {20}},\ \bibinfo
		{pages} {045804} (\bibinfo {year} {2018})}\BibitemShut {NoStop}%
	\bibitem [{\citenamefont {Chung}\ \emph {et~al.}(2019)\citenamefont {Chung},
		\citenamefont {{Martinez-Garaot}}, \citenamefont {Chen}, \citenamefont
		{Muga},\ and\ \citenamefont {Tseng}}]{Chung_2019}%
	\BibitemOpen
	\bibfield  {author} {\bibinfo {author} {\bibfnamefont {H.-C.}\ \bibnamefont
			{Chung}}, \bibinfo {author} {\bibfnamefont {S.}~\bibnamefont
			{{Martinez-Garaot}}}, \bibinfo {author} {\bibfnamefont {X.}~\bibnamefont
			{Chen}}, \bibinfo {author} {\bibfnamefont {J.~G.}\ \bibnamefont {Muga}}, \
		and\ \bibinfo {author} {\bibfnamefont {S.-Y.}\ \bibnamefont {Tseng}},\ }\href
	{\doibase 10.1209/0295-5075/127/34001} {\bibfield  {journal} {\bibinfo
			{journal} {EPL}\ }\textbf {\bibinfo {volume} {127}},\
		\bibinfo {pages} {34001} (\bibinfo {year} {2019})}\BibitemShut {NoStop}%
	\bibitem [{\citenamefont {Dou}\ \emph {et~al.}(2020)\citenamefont {Dou},
		\citenamefont {Yan}, \citenamefont {Liu}, \citenamefont {Wang},\ and\
		\citenamefont {Shu}}]{Dou_2020}%
	\BibitemOpen
	\bibfield  {author} {\bibinfo {author} {\bibfnamefont {F.-Q.}\ \bibnamefont
			{Dou}}, \bibinfo {author} {\bibfnamefont {Z.-M.}\ \bibnamefont {Yan}},
		\bibinfo {author} {\bibfnamefont {X.-Q.}\ \bibnamefont {Liu}}, \bibinfo
		{author} {\bibfnamefont {W.-Y.}\ \bibnamefont {Wang}}, \ and\ \bibinfo
		{author} {\bibfnamefont {C.-C.}\ \bibnamefont {Shu}},\ }\href {\doibase
		10.1016/j.ijleo.2020.164516} {\bibfield  {journal} {\bibinfo  {journal}
			{Optik}\ }\textbf {\bibinfo {volume} {210}},\ \bibinfo {pages} {164516}
		(\bibinfo {year} {2020})}\BibitemShut {NoStop}%
	\bibitem [{\citenamefont {Moiseyev}(2011)}]{Book_moiseyev_2011}%
	\BibitemOpen
	\bibfield  {author} {\bibinfo {author} {\bibfnamefont {N.}~\bibnamefont
			{Moiseyev}},\ }\href {\doibase 10.1017/CBO9780511976186} {\emph {\bibinfo
			{title} {Non-Hermitian Quantum Mechanics}}}\ (\bibinfo  {publisher}
	{{Cambridge University Press}},\ \bibinfo {address} {{Cambridge, UK}},\ \bibinfo
	{year} {2011})\BibitemShut {NoStop}%
	\bibitem [{\citenamefont {Uzdin}\ and\ \citenamefont
		{Moiseyev}(2012)}]{Uzdin_2012}%
	\BibitemOpen
	\bibfield  {author} {\bibinfo {author} {\bibfnamefont {R.}~\bibnamefont
			{Uzdin}}\ and\ \bibinfo {author} {\bibfnamefont {N.}~\bibnamefont
			{Moiseyev}},\ }\href {\doibase 10.1088/1751-8113/45/44/444033} {\bibfield
		{journal} {\bibinfo  {journal} {J. Phys. A}\ }\textbf {\bibinfo {volume}
			{45}},\ \bibinfo {pages} {444033} (\bibinfo {year} {2012})}\BibitemShut
	{NoStop}%
	\bibitem [{\citenamefont {Vitanov}\ and\ \citenamefont
		{Stenholm}(1997)}]{Vitanov_PRA97}%
	\BibitemOpen
	\bibfield  {author} {\bibinfo {author} {\bibfnamefont {N.~V.}\ \bibnamefont
			{Vitanov}}\ and\ \bibinfo {author} {\bibfnamefont {S.}~\bibnamefont
			{Stenholm}},\ }\href {\doibase 10.1103/PhysRevA.56.1463} {\bibfield
		{journal} {\bibinfo  {journal} {Phys. Rev. A}\ }\textbf {\bibinfo {volume}
			{56}},\ \bibinfo {pages} {1463} (\bibinfo {year} {1997})}\BibitemShut
	{NoStop}%
	\bibitem [{\citenamefont {Fang}\ and\ \citenamefont {Sun}(2015)}]{Fang2015}%
	\BibitemOpen
	\bibfield  {author} {\bibinfo {author} {\bibfnamefont {Y.}~\bibnamefont
			{Fang}}\ and\ \bibinfo {author} {\bibfnamefont {M.}~\bibnamefont {Sun}},\
	}\href@noop {} {\bibfield  {journal} {\bibinfo  {journal} {Light Sci. Appl.}\
		}\textbf {\bibinfo {volume} {4}},\ \bibinfo {pages} {e294} (\bibinfo {year}
		{2015})}\BibitemShut {NoStop}%
	\bibitem [{\citenamefont {Guo}\ \emph {et~al.}(2013)\citenamefont {Guo},
		\citenamefont {Ma}, \citenamefont {Wang},\ and\ \citenamefont
		{Tong}}]{Guo2013}%
	\BibitemOpen
	\bibfield  {author} {\bibinfo {author} {\bibfnamefont {X.}~\bibnamefont
			{Guo}}, \bibinfo {author} {\bibfnamefont {Y.}~\bibnamefont {Ma}}, \bibinfo
		{author} {\bibfnamefont {Y.}~\bibnamefont {Wang}}, \ and\ \bibinfo {author}
		{\bibfnamefont {L.}~\bibnamefont {Tong}},\ }\href {\doibase
		10.1002/lpor.201200067} {\bibfield  {journal} {\bibinfo  {journal} {Laser
				{$\&$} Photon. Rev.}\ }\textbf {\bibinfo {volume} {7}},\ \bibinfo {pages}
		{855} (\bibinfo {year} {2013})}\BibitemShut {NoStop}%
	\bibitem [{\citenamefont {Yariv}(1973)}]{yariv_coupled-mode_1973}%
	\BibitemOpen
	\bibfield  {author} {\bibinfo {author} {\bibfnamefont {A.}~\bibnamefont
			{Yariv}},\ }\href {\doibase 10.1109/JQE.1973.1077767} {\bibfield  {journal}
		{\bibinfo  {journal} {IEEE J. Quantum Electron.}\ }\textbf {\bibinfo {volume}
			{9}},\ \bibinfo {pages} {919} (\bibinfo {year} {1973})}\BibitemShut {NoStop}%
	\bibitem [{\citenamefont {Ciret}\ \emph {et~al.}(2013)\citenamefont {Ciret},
		\citenamefont {Alonzo}, \citenamefont {Coda}, \citenamefont {Rangelov},\ and\
		\citenamefont {Montemezzani}}]{Ciret_EIT_2013}%
	\BibitemOpen
	\bibfield  {author} {\bibinfo {author} {\bibfnamefont {C.}~\bibnamefont
			{Ciret}}, \bibinfo {author} {\bibfnamefont {M.}~\bibnamefont {Alonzo}},
		\bibinfo {author} {\bibfnamefont {V.}~\bibnamefont {Coda}}, \bibinfo {author}
		{\bibfnamefont {A.~A.}\ \bibnamefont {Rangelov}}, \ and\ \bibinfo {author}
		{\bibfnamefont {G.}~\bibnamefont {Montemezzani}},\ }\href {\doibase
		10.1103/PhysRevA.88.013840} {\bibfield  {journal} {\bibinfo  {journal} {Phys.
				Rev. A}\ }\textbf {\bibinfo {volume} {88}},\ \bibinfo {pages} {013840}
		(\bibinfo {year} {2013})}\BibitemShut {NoStop}%
	\bibitem [{\citenamefont {Shore}\ and\ \citenamefont {{N. V.
				Vitanov}}(2006)}]{Shore_2006}%
	\BibitemOpen
	\bibfield  {author} {\bibinfo {author} {\bibfnamefont {B.~W.}\ \bibnamefont
			{Shore}}\ and\ \bibinfo {author} {\bibnamefont {{N. V. Vitanov}}},\ }\href
	{\doibase 10.1080/00107510601181159} {\bibfield  {journal} {\bibinfo
			{journal} {Contemp. Phys.}\ }\textbf {\bibinfo {volume} {47}},\ \bibinfo
		{pages} {341} (\bibinfo {year} {2006})}\BibitemShut {NoStop}%
	\bibitem [{\citenamefont {Roey}\ \emph {et~al.}(1981)\citenamefont {Roey},
		\citenamefont {{van der Donk}},\ and\ \citenamefont {Lagasse}}]{VanRoey_81}%
	\BibitemOpen
	\bibfield  {author} {\bibinfo {author} {\bibfnamefont {J.~V.}\ \bibnamefont
			{Roey}}, \bibinfo {author} {\bibfnamefont {J.}~\bibnamefont {{van der
					Donk}}}, \ and\ \bibinfo {author} {\bibfnamefont {P.~E.}\ \bibnamefont
			{Lagasse}},\ }\href@noop {} {\bibfield  {journal} {\bibinfo  {journal} {J.
				Opt. Soc. Am.}\ }\textbf {\bibinfo {volume} {71}},\ \bibinfo {pages} {803}
		(\bibinfo {year} {1981})}\BibitemShut {NoStop}%
	\bibitem [{\citenamefont {Kawano}\ and\ \citenamefont
		{Kitoh}(2001)}]{Book_Kawano}%
	\BibitemOpen
	\bibfield  {author} {\bibinfo {author} {\bibfnamefont {K.}~\bibnamefont
			{Kawano}}\ and\ \bibinfo {author} {\bibfnamefont {T.}~\bibnamefont {Kitoh}},\
	}\href {\doibase 10.1017/CBO9780511976186} {\emph {\bibinfo {title}
			{Introduction to Optical Waveguide Analysis: {{Solving Maxwell}}'s Equations
				and the {{Schr\"odinger}} Equation}}}\ (\bibinfo  {publisher} {{John Wiley
			{$\&$} Sons}},\ \bibinfo {address} {{New York}},\ \bibinfo {year}
	{2001})\BibitemShut {NoStop}%
	\bibitem [{\citenamefont {Malitson}(1965)}]{Malitson65}%
	\BibitemOpen
	\bibfield  {author} {\bibinfo {author} {\bibfnamefont {I.~H.}\ \bibnamefont
			{Malitson}},\ }\href {\doibase 10.1364/JOSA.55.001205} {\bibfield  {journal}
		{\bibinfo  {journal} {J. Opt. Soc. Am.}\ }\textbf {\bibinfo {volume} {55}},\
		\bibinfo {pages} {1205} (\bibinfo {year} {1965})}\BibitemShut {NoStop}%
	\bibitem [{\citenamefont {Naresh}\ and\ \citenamefont
		{Buddhudu}(2015)}]{Naresh2015}%
	\BibitemOpen
	\bibfield  {author} {\bibinfo {author} {\bibfnamefont {V.}~\bibnamefont
			{Naresh}}\ and\ \bibinfo {author} {\bibfnamefont {S.}~\bibnamefont
			{Buddhudu}},\ }\href {\doibase doi:10.13036/17533562.56.5.255} {\bibfield
		{journal} {\bibinfo  {journal} {Phys. Chem. Glasses: Eur. J. Glass Sci.
				Technol. B}\ }\textbf {\bibinfo {volume} {56}},\ \bibinfo {pages} {255}
		(\bibinfo {year} {2015})}\BibitemShut {NoStop}%
	\bibitem [{\citenamefont {Tikhomirov}\ \emph {et~al.}(2002)\citenamefont
		{Tikhomirov}, \citenamefont {Furniss}, \citenamefont {Seddon}, \citenamefont
		{Ferrari},\ and\ \citenamefont {Rolli}}]{Tikhomirov2002}%
	\BibitemOpen
	\bibfield  {author} {\bibinfo {author} {\bibfnamefont {V.~K.}\ \bibnamefont
			{Tikhomirov}}, \bibinfo {author} {\bibfnamefont {D.}~\bibnamefont {Furniss}},
		\bibinfo {author} {\bibfnamefont {A.~B.}\ \bibnamefont {Seddon}}, \bibinfo
		{author} {\bibfnamefont {M.}~\bibnamefont {Ferrari}}, \ and\ \bibinfo
		{author} {\bibfnamefont {R.}~\bibnamefont {Rolli}},\ }\href@noop {}
	{\bibfield  {journal} {\bibinfo  {journal} {J. Mater. Sci. Lett.}\ }\textbf
		{\bibinfo {volume} {21}},\ \bibinfo {pages} {293} (\bibinfo {year}
		{2002})}\BibitemShut {NoStop}%
	\bibitem [{\citenamefont {Rolli}\ \emph {et~al.}(2003)\citenamefont {Rolli},
		\citenamefont {Montagna}, \citenamefont {Chaussedent}, \citenamefont
		{Monteil}, \citenamefont {Tikhomirov},\ and\ \citenamefont
		{Ferrari}}]{Rolli2003}%
	\BibitemOpen
	\bibfield  {author} {\bibinfo {author} {\bibfnamefont {R.}~\bibnamefont
			{Rolli}}, \bibinfo {author} {\bibfnamefont {M.}~\bibnamefont {Montagna}},
		\bibinfo {author} {\bibfnamefont {S.}~\bibnamefont {Chaussedent}}, \bibinfo
		{author} {\bibfnamefont {A.}~\bibnamefont {Monteil}}, \bibinfo {author}
		{\bibfnamefont {V.}~\bibnamefont {Tikhomirov}}, \ and\ \bibinfo {author}
		{\bibfnamefont {M.}~\bibnamefont {Ferrari}},\ }\href {\doibase
		10.1016/S0925-3467(02)00092-7} {\bibfield  {journal} {\bibinfo  {journal}
			{Opt. Mater.}\ }\textbf {\bibinfo {volume} {21}},\ \bibinfo {pages} {743}
		(\bibinfo {year} {2003})}\BibitemShut {NoStop}%
	\bibitem [{\citenamefont {Sohler}\ \emph {et~al.}(2005)\citenamefont {Sohler},
		\citenamefont {Das}, \citenamefont {Dey}, \citenamefont {Reza}, \citenamefont
		{Suche},\ and\ \citenamefont {Ricken}}]{Sohler2005}%
	\BibitemOpen
	\bibfield  {author} {\bibinfo {author} {\bibfnamefont {W.}~\bibnamefont
			{Sohler}}, \bibinfo {author} {\bibfnamefont {B.~K.}\ \bibnamefont {Das}},
		\bibinfo {author} {\bibfnamefont {D.}~\bibnamefont {Dey}}, \bibinfo {author}
		{\bibfnamefont {S.}~\bibnamefont {Reza}}, \bibinfo {author} {\bibfnamefont
			{H.}~\bibnamefont {Suche}}, \ and\ \bibinfo {author} {\bibfnamefont
			{R.}~\bibnamefont {Ricken}},\ }\href@noop {} {\bibfield  {journal} {\bibinfo
			{journal} {IEICE Trans. Electron. C}\ }\textbf {\bibinfo {volume} {E88}},\
		\bibinfo {pages} {990} (\bibinfo {year} {2005})}\BibitemShut {NoStop}%
	\bibitem [{\citenamefont {Mattarelli}\ \emph {et~al.}(2006)\citenamefont
		{Mattarelli}, \citenamefont {Sebastiani}, \citenamefont {Spirkova},
		\citenamefont {Berneschi}, \citenamefont {Brenci}, \citenamefont {Calzolai},
		\citenamefont {Chiasera}, \citenamefont {Ferrari}, \citenamefont {Montagna},
		\citenamefont {Conti}, \citenamefont {Pelli},\ and\ \citenamefont
		{Righini}}]{Mattarelli2006}%
	\BibitemOpen
	\bibfield  {author} {\bibinfo {author} {\bibfnamefont {M.}~\bibnamefont
			{Mattarelli}}, \bibinfo {author} {\bibfnamefont {S.}~\bibnamefont
			{Sebastiani}}, \bibinfo {author} {\bibfnamefont {J.}~\bibnamefont
			{Spirkova}}, \bibinfo {author} {\bibfnamefont {S.}~\bibnamefont {Berneschi}},
		\bibinfo {author} {\bibfnamefont {M.}~\bibnamefont {Brenci}}, \bibinfo
		{author} {\bibfnamefont {R.}~\bibnamefont {Calzolai}}, \bibinfo {author}
		{\bibfnamefont {A.}~\bibnamefont {Chiasera}}, \bibinfo {author}
		{\bibfnamefont {M.}~\bibnamefont {Ferrari}}, \bibinfo {author} {\bibfnamefont
			{M.}~\bibnamefont {Montagna}}, \bibinfo {author} {\bibfnamefont {G.~N.}\
			\bibnamefont {Conti}}, \bibinfo {author} {\bibfnamefont {S.}~\bibnamefont
			{Pelli}}, \ and\ \bibinfo {author} {\bibfnamefont {G.}~\bibnamefont
			{Righini}},\ }\href {\doibase 10.1016/j.optmat.2006.01.030} {\bibfield
		{journal} {\bibinfo  {journal} {Opt. Mater.}\ }\textbf {\bibinfo {volume}
			{28}},\ \bibinfo {pages} {1292} (\bibinfo {year} {2006})}\BibitemShut
	{NoStop}%
	\bibitem [{\citenamefont {Yan}\ \emph {et~al.}(1997)\citenamefont {Yan},
		\citenamefont {Faber}, \citenamefont {{de Waal}}, \citenamefont {Kik},\ and\
		\citenamefont {Polman}}]{Yan1997}%
	\BibitemOpen
	\bibfield  {author} {\bibinfo {author} {\bibfnamefont {Y.~C.}\ \bibnamefont
			{Yan}}, \bibinfo {author} {\bibfnamefont {A.~J.}\ \bibnamefont {Faber}},
		\bibinfo {author} {\bibfnamefont {H.}~\bibnamefont {{de Waal}}}, \bibinfo
		{author} {\bibfnamefont {P.~G.}\ \bibnamefont {Kik}}, \ and\ \bibinfo
		{author} {\bibfnamefont {A.}~\bibnamefont {Polman}},\ }\href {\doibase
		10.1063/1.120216} {\bibfield  {journal} {\bibinfo  {journal} {Appl. Phys.
				Lett.}\ }\textbf {\bibinfo {volume} {71}},\ \bibinfo {pages} {2922} (\bibinfo
		{year} {1997})}\BibitemShut {NoStop}%
	\bibitem [{\citenamefont {Wetter}\ \emph {et~al.}(2019)\citenamefont {Wetter},
		\citenamefont {{Silverio da Silva}}, \citenamefont {Reyes Pires~Kassab},\
		and\ \citenamefont {{Jimenez-Villar}}}]{Wetter2019}%
	\BibitemOpen
	\bibfield  {author} {\bibinfo {author} {\bibfnamefont {N.~U.}\ \bibnamefont
			{Wetter}}, \bibinfo {author} {\bibfnamefont {D.}~\bibnamefont {{Silverio da
					Silva}}}, \bibinfo {author} {\bibfnamefont {L.}~\bibnamefont {Reyes
				Pires~Kassab}}, \ and\ \bibinfo {author} {\bibfnamefont {E.}~\bibnamefont
			{{Jimenez-Villar}}},\ }\href {\doibase 10.1016/j.jallcom.2019.04.141}
	{\bibfield  {journal} {\bibinfo  {journal} {J. Alloys Compd.}\ }\textbf
		{\bibinfo {volume} {794}},\ \bibinfo {pages} {120} (\bibinfo {year}
		{2019})}\BibitemShut {NoStop}%
	\bibitem [{\citenamefont {Slooff}\ \emph {et~al.}(2002)\citenamefont {Slooff},
		\citenamefont {{van Blaaderen}}, \citenamefont {Polman}, \citenamefont
		{Hebbink}, \citenamefont {Klink}, \citenamefont {Van~Veggel}, \citenamefont
		{Reinhoudt},\ and\ \citenamefont {Hofstraat}}]{Sloof2002}%
	\BibitemOpen
	\bibfield  {author} {\bibinfo {author} {\bibfnamefont {L.~H.}\ \bibnamefont
			{Slooff}}, \bibinfo {author} {\bibfnamefont {A.}~\bibnamefont {{van
					Blaaderen}}}, \bibinfo {author} {\bibfnamefont {A.}~\bibnamefont {Polman}},
		\bibinfo {author} {\bibfnamefont {G.~A.}\ \bibnamefont {Hebbink}}, \bibinfo
		{author} {\bibfnamefont {S.~I.}\ \bibnamefont {Klink}}, \bibinfo {author}
		{\bibfnamefont {F.~C. J.~M.}\ \bibnamefont {Van~Veggel}}, \bibinfo {author}
		{\bibfnamefont {D.~N.}\ \bibnamefont {Reinhoudt}}, \ and\ \bibinfo {author}
		{\bibfnamefont {J.~W.}\ \bibnamefont {Hofstraat}},\ }\href {\doibase
		10.1063/1.1454190} {\bibfield  {journal} {\bibinfo  {journal} {Journal of
				Applied Physics}\ }\textbf {\bibinfo {volume} {91}},\ \bibinfo {pages} {3955}
		(\bibinfo {year} {2002})}\BibitemShut {NoStop}%
	\bibitem [{\citenamefont {Ziarko}\ \emph {et~al.}(2019)\citenamefont {Ziarko},
		\citenamefont {Bamiedakis}, \citenamefont {{Kumi-Barimah}}, \citenamefont
		{Jose}, \citenamefont {Penty},\ and\ \citenamefont {White}}]{Ziarko2019}%
	\BibitemOpen
	\bibfield  {author} {\bibinfo {author} {\bibfnamefont {M.}~\bibnamefont
			{Ziarko}}, \bibinfo {author} {\bibfnamefont {N.}~\bibnamefont {Bamiedakis}},
		\bibinfo {author} {\bibfnamefont {E.}~\bibnamefont {{Kumi-Barimah}}},
		\bibinfo {author} {\bibfnamefont {G.}~\bibnamefont {Jose}}, \bibinfo {author}
		{\bibfnamefont {R.~V.}\ \bibnamefont {Penty}}, \ and\ \bibinfo {author}
		{\bibfnamefont {I.~H.}\ \bibnamefont {White}},\ }\href {\doibase
		10.1117/12.2509983} {\bibfield  {journal} {\bibinfo  {journal} {Proc. SPIE}\
		}\textbf {\bibinfo {volume} {10924}},\ \bibinfo {pages} {1092403 } (\bibinfo {year}
		{2019})}\BibitemShut {NoStop}%
	\bibitem [{\citenamefont {Catchpole}\ and\ \citenamefont
		{Pillai}(2006)}]{Catchpole2006}%
	\BibitemOpen
	\bibfield  {author} {\bibinfo {author} {\bibfnamefont {K.~R.}\ \bibnamefont
			{Catchpole}}\ and\ \bibinfo {author} {\bibfnamefont {S.}~\bibnamefont
			{Pillai}},\ }\href {\doibase 10.1063/1.2226334} {\bibfield  {journal}
		{\bibinfo  {journal} {J. Appl. Phys.}\ }\textbf {\bibinfo {volume} {100}},\
		\bibinfo {pages} {044504} (\bibinfo {year} {2006})}\BibitemShut {NoStop}%
	\bibitem [{\citenamefont {Biesenthal}\ \emph {et~al.}(2019)\citenamefont
		{Biesenthal}, \citenamefont {Kremer}, \citenamefont {Heinrich},\ and\
		\citenamefont {Szameit}}]{BiesenthalPRL2019}%
	\BibitemOpen
	\bibfield  {author} {\bibinfo {author} {\bibfnamefont {T.}~\bibnamefont
			{Biesenthal}}, \bibinfo {author} {\bibfnamefont {M.}~\bibnamefont {Kremer}},
		\bibinfo {author} {\bibfnamefont {M.}~\bibnamefont {Heinrich}}, \ and\
		\bibinfo {author} {\bibfnamefont {A.}~\bibnamefont {Szameit}},\ }\href
	{\doibase 10.1103/PhysRevLett.123.183601} {\bibfield  {journal} {\bibinfo
			{journal} {Phys. Rev. Lett.}\ }\textbf {\bibinfo {volume} {123}},\ \bibinfo
		{pages} {183601} (\bibinfo {year} {2019})}\BibitemShut {NoStop}%
	\bibitem [{\citenamefont {Li}\ \emph {et~al.}(2010)\citenamefont {Li},
		\citenamefont {Song}, \citenamefont {Zhou}, \citenamefont {Su},\ and\
		\citenamefont {Qiu}}]{Li2010}%
	\BibitemOpen
	\bibfield  {author} {\bibinfo {author} {\bibfnamefont {Q.}~\bibnamefont
			{Li}}, \bibinfo {author} {\bibfnamefont {Y.}~\bibnamefont {Song}}, \bibinfo
		{author} {\bibfnamefont {G.}~\bibnamefont {Zhou}}, \bibinfo {author}
		{\bibfnamefont {Y.}~\bibnamefont {Su}}, \ and\ \bibinfo {author}
		{\bibfnamefont {M.}~\bibnamefont {Qiu}},\ }\href {\doibase
		10.1364/OL.35.003153} {\bibfield  {journal} {\bibinfo  {journal} {Opt.
				Lett.}\ }\textbf {\bibinfo {volume} {35}},\ \bibinfo {pages} {3153} (\bibinfo
		{year} {2010})}\BibitemShut {NoStop}%
	\bibitem [{\citenamefont {Briggs}\ \emph {et~al.}(2010)\citenamefont {Briggs},
		\citenamefont {Grandidier}, \citenamefont {Burgos}, \citenamefont
		{Feigenbaum},\ and\ \citenamefont {Atwater}}]{Briggs2010}%
	\BibitemOpen
	\bibfield  {author} {\bibinfo {author} {\bibfnamefont {R.~M.}\ \bibnamefont
			{Briggs}}, \bibinfo {author} {\bibfnamefont {J.}~\bibnamefont {Grandidier}},
		\bibinfo {author} {\bibfnamefont {S.~P.}\ \bibnamefont {Burgos}}, \bibinfo
		{author} {\bibfnamefont {E.}~\bibnamefont {Feigenbaum}}, \ and\ \bibinfo
		{author} {\bibfnamefont {H.~A.}\ \bibnamefont {Atwater}},\ }\href {\doibase
		10.1021/nl1024529} {\bibfield  {journal} {\bibinfo  {journal} {Nano Lett.}\
		}\textbf {\bibinfo {volume} {10}},\ \bibinfo {pages} {4851} (\bibinfo {year}
		{2010})}\BibitemShut {NoStop}%
	\bibitem [{\citenamefont {Lou}\ \emph {et~al.}(2012)\citenamefont {Lou},
		\citenamefont {Dai},\ and\ \citenamefont {Wosinski}}]{Lou2012}%
	\BibitemOpen
	\bibfield  {author} {\bibinfo {author} {\bibfnamefont {F.}~\bibnamefont
			{Lou}}, \bibinfo {author} {\bibfnamefont {D.}~\bibnamefont {Dai}}, \ and\
		\bibinfo {author} {\bibfnamefont {L.}~\bibnamefont {Wosinski}},\ }\href
	{\doibase 10.1364/OL.37.003372} {\bibfield  {journal} {\bibinfo  {journal}
			{Opt. Lett.}\ }\textbf {\bibinfo {volume} {37}},\ \bibinfo {pages} {3372}
		(\bibinfo {year} {2012})}\BibitemShut {NoStop}%
	\bibitem [{\citenamefont {Chee}\ \emph {et~al.}(2012)\citenamefont {Chee},
		\citenamefont {Zhu},\ and\ \citenamefont {Lo}}]{Chee2012}%
	\BibitemOpen
	\bibfield  {author} {\bibinfo {author} {\bibfnamefont {J.}~\bibnamefont
			{Chee}}, \bibinfo {author} {\bibfnamefont {S.}~\bibnamefont {Zhu}}, \ and\
		\bibinfo {author} {\bibfnamefont {G.~Q.}\ \bibnamefont {Lo}},\ }\href
	{\doibase 10.1364/OE.20.025345} {\bibfield  {journal} {\bibinfo  {journal}
			{Opt. Express}\ }\textbf {\bibinfo {volume} {20}},\ \bibinfo {pages} {25345}
		(\bibinfo {year} {2012})}\BibitemShut {NoStop}%
	\bibitem [{\citenamefont {Guan}\ \emph {et~al.}(2013)\citenamefont {Guan},
		\citenamefont {Wu}, \citenamefont {Shi}, \citenamefont {Wosinski},\ and\
		\citenamefont {Dai}}]{Guan2013}%
	\BibitemOpen
	\bibfield  {author} {\bibinfo {author} {\bibfnamefont {X.}~\bibnamefont
			{Guan}}, \bibinfo {author} {\bibfnamefont {H.}~\bibnamefont {Wu}}, \bibinfo
		{author} {\bibfnamefont {Y.}~\bibnamefont {Shi}}, \bibinfo {author}
		{\bibfnamefont {L.}~\bibnamefont {Wosinski}}, \ and\ \bibinfo {author}
		{\bibfnamefont {D.}~\bibnamefont {Dai}},\ }\href {\doibase
		10.1364/OL.38.003005} {\bibfield  {journal} {\bibinfo  {journal} {Opt.
				Lett.}\ }\textbf {\bibinfo {volume} {38}},\ \bibinfo {pages} {3005} (\bibinfo
		{year} {2013})}\BibitemShut {NoStop}%
\end{thebibliography}


%


\end{document}